\documentclass[prb,aps,twocolumn,showpacs,superscriptaddress,amsmath,amssymb]{revtex4}
\usepackage{colordvi}
\usepackage{psfig}
\begin{document}

\title{Remote-control spin filtering through a $T$-type structure}
\author{X. Y. Feng}
\affiliation{Hefei National Laboratory for Physical Sciences at
  Microscale, University of Science and
Technology of China, Hefei, Anhui, 230026, China}
\affiliation{Department of Physics, University of Science and
Technology of China, Hefei, Anhui, 230026, China}
\altaffiliation{Mailing Address.}
\author{J. H. Jiang}
\affiliation{Department of Physics, University of Science and
Technology of China, Hefei, Anhui, 230026, China}
\author{M. Q. Weng}
\thanks{Author to whom correspondence should be addressed}%
\email{weng@ustc.edu.cn}%
\affiliation{Department of Physics, University of Science and
Technology of China, Hefei, Anhui, 230026, China}

\date{\today}

\begin{abstract}
We propose a spin filter scheme using a $T$-stub waveguide. By applying
a moderate magnetic field at the tip of the sidearm, this device can
produce both large electric and spin current. The direction,
polarization of the output spin current can be further adjusted
electronically by a remote gate which tunes the length of the
sidearm. The device is robust against the disorder.
\end{abstract}


\maketitle

Spintronic devices have many advantages over the traditional
electronic devices such as higher operation speed, lower power
consumption.\cite{datta,wolf,das} Tremendous efforts have been
devoted in the past few years to overcome the fundamental obstacles in
the realization of spintronic devices, such as spin generation,
control and detection. Direct injection of spin current from
ferromagnetic metal or semiconductor was first proposed to realize
the spin injection.\cite{fiederling_1999,oestreich_1999}
Despite the great efforts, the highest spin
injection rate reported is about $90\%$\cite{fiederling_1999}
for p-typed and $57\%$\cite{jiang} for n-typed semiconductor.
A great number of schemes of spin filter have been
proposed to produce highly polarized spin current using various of
structures such as electronic waveguide,\cite{zhai,zhou_apl_2004}
double-bend structure,\cite{zhou_apl_2005}
Aharonov-Bohm (AB) ring,\cite{frustaglia}
resonant tunneling diode,\cite{koga} 
quantum dot,\cite{recher,karkka,folk}
ferromagnetic electrode.\cite{grundler_2001_prb}
More recently, the spin filter for the hole system was brought to
public.\cite{wu_2005_prb}
Dynamical spin generation in semiconductor using
oscillation field is also
proposed to avoid the difficulty of spin
injection through interface.\cite{cheng_2005_apl,zhang}

\begin{figure}[htb]
\vskip-0.1cm
\begin{center}
  \psfig{figure=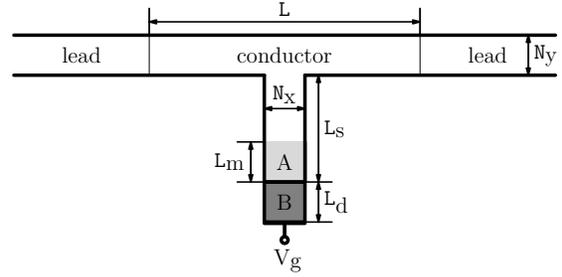,width=0.85\columnwidth}
\end{center}
\caption{Schematic of remote-control spin filter in T-stub
  waveguide. The modulating field is applied only at the tip of the
  sidearm (area A and B).
}
\label{fig1}
\end{figure}

In this paper, we propose a spin filter scheme which can produce high
spin polarization (SP) and also provides remote control of spin
current magnetically as well as electronically.  Our scheme
is a localized magnetic field modulated 2-dimensional (2D) T-stub
waveguide as shown in
Fig.~\ref{fig1}.  The waveguide is composed of a longitudinal
conductor with length $L$, width $N_y$ and a sidearm of width $N_x$
attached to the center of the conductor.  The effective conducting
length of sidearm $L_s$ can be controlled by the remote gate voltage
$V_g$ which changes $L_d$, the length of depletion area (area B in
Fig.~\ref{fig1}.  The far edge of
sidearm (gray areas A and B in the Fig.~\ref{fig1}) is modulated by an
applied magnetic field which gives Zeeman splitting of $2 V_0$.
T-stub geometry has long been proposed as quantum modulated transistor
(QMT).\cite{sols_jap_1989,aihara_apl_1993} The conductance of this
kind of devices is determined by the quantum interference effect
between different Feynman paths and oscillates with the Fermi energy.
When a modulate magnetic field is applied, the electrons of different
spins will have different phase shifts transpassing
through the modulated area and therefore have different
conductances. In this way, the non-spin-polarized electron current
pass through this device will produce SP. The spin current can also be
controlled electronically by the remote gate by adjusting the
effective sidearm length $L_s$. 
It is worth noting that our spin
filter is different from other implementations in the fact that
our device can be {\em remotely} controlled by both magnetical and
electronic methods while the other ones can only be controlled locally.
Moreover our device has more energy windows to generate
high polarized spin current than the device made by
a single quantum dot.\cite{folk} 

We describe the T-stub geometry by the tight-binding Hamiltonian with
nearest-neighbor approximation:
\begin{eqnarray}
  H&=&\sum_{l,m,\sigma} \varepsilon_{l,m,\sigma} c^\dagger_{l,m,\sigma}
  c_{l,m,\sigma} +
  t_0 \sum_{l,m,\sigma} (c^\dagger_{l+1,m,\sigma}c_{l,m,\sigma} \nonumber \\
  &&+ c^\dagger_{l,m+1,\sigma}c_{l,m,\sigma} +h.c.)\
\end{eqnarray}
in which $l$ and $m$ denote the ``lattice'' site index along the $x$-
and $y$-axis
respectively. The on-site energy
$\varepsilon_{l,m,\sigma}=\varepsilon_0+\sigma V_0$  when $(l,m)$
locates in the modulated regime (the gray area in the Fig.~\ref{fig1})
and $\varepsilon_0$ otherwise. $\varepsilon_0=-4t_0$ and
$t_0=-\hbar/(2m^{\ast}a^2)$ is the hopping energy with $m^{\ast}$ and
$a$ standing for the effective mass and the ``lattice'' constant
respectively. $\sigma V_0$ is the Zeeman energy of spin $\sigma$ in
the modulate magnetic field and $\sigma=\pm 1$ for spin-up and -down
electrons respectively.

The two-terminal spin-dependent conductance is obtained by using
Landauer-B\"uttiker\cite{buttiker} formula
$G^{\sigma \sigma^\prime}(E)=(e^2/h)\mathtt{Tr}[\Gamma^{\sigma}_{1}
G^{\sigma\sigma^\prime+}_{1L}(E)\Gamma^{\sigma^\prime}_{N}G^{\sigma
^\prime\sigma -}
_{L1}(E)]$
with $\Gamma^{\sigma}_{1/N}$ representing the self-energy function
for the isolated left/right ideal leads.\cite{datta_95}
$G^{\sigma\sigma^\prime+}_{1L}(E)$ and
$G^{\sigma\sigma^\prime-}_{L1}(E)$ are the full retarded and advanced Green
functions for the conductor which have taken account the effect of the
leads. $\mathtt{Tr}$ stands for the trace over the $y$-axis. The spin
dependent current is given by $I_{\sigma}=\int_E^{E+\Delta}
G^{\sigma\sigma}(\varepsilon) d\varepsilon$ for energy window
$[E,E+\Delta]$.

\begin{figure}[htb]
\vskip-0.3cm
\begin{center}
  \psfig{figure=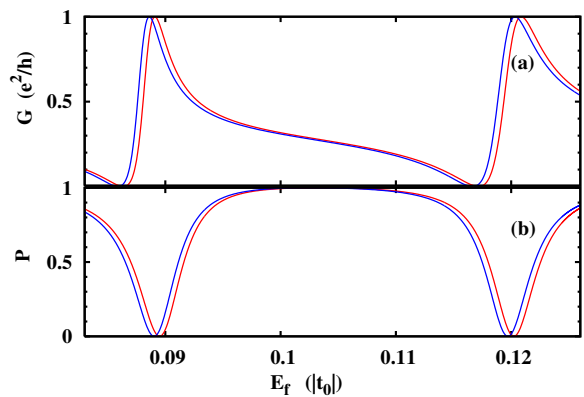,width=0.95\columnwidth}
\end{center}
\vskip-0.3cm
  \caption{(Color online)
Spin dependent conductances
$G^{\uparrow\uparrow}$ (red curve) and
$G^{\downarrow\downarrow}$ (blue curve) vs the incident energy
of the electrons for quasi-one-dimensional (a) and two-dimensional (b).
}
\label{fig2}
\end{figure}

We first study the analytically solvable
transport problem in quasi-one-dimensional ($N_x=N_y=1$) T-stub
geometry system.
By solving the time independent Schr\"odinger equation,\cite{xiong}
the transmission coefficient in this system
can be written as
\begin{eqnarray}
&&t_{\sigma}=2i G^{\sigma\sigma}_{0,0}(E)t_0^3 \sin(ka)/\{
[\varepsilon_1-E+t_0e^{ika}+G^{\sigma\sigma}_{0,0}(E)t_0^2]
\nonumber \\
&&\times
[\varepsilon_{-1}-E+t_0e^{ika}+G^{\sigma\sigma}_{0,0}(E)t_0^2]
-[G^{\sigma\sigma}_{0,0}(E)t_0^2]^2\},
\end{eqnarray}
in which $G^{\sigma\sigma^{\prime}}_{l,m}(E)
=(E-\hat{H})^{-1}_{l\sigma,m\sigma^{\prime}}$ and $\hat{H}$ are the
Green function and the Hamiltonian of sidearm respectively.
$E=2|t_0|(1-\cos(k\,a))$ is the energy of electron with wave-vector $k$.
The corresponding conductance is
$G^{\sigma}=(e^2/h)|t_{\sigma}|^2$.
Without the modulation, the transmission coefficient drops to zero at
the anti-resonance points $2|t_0|[1-\cos (n\pi/(L_s+1))] (n=1,2,\cdots)$
and reaches its maximum quickly at resonance points $2|t_0|[1-\cos
(n\pi/L_s)] (n=1,2,\cdots)$.
When the modulate field turns on, the the (anti-)resonance points of
spin-up and -down shift apart.
In Fig.~\ref{fig2}(a) we plot spin resolved conductance as a function
of the electron energy for a typical quasi-one-dimensional T-stub
device with sidearm length $L_s=63\,a$, magnetic field
$V_0=0.001\,|t_0|$ and modulation length $L_m=10\,a$.
Since a pair of anti-resonance and
resonance points are very close to each other when the sidearm is
relatively long, it is possible to choose the system parameters so that the
resonance point of spin-up (-down) electron matches the anti-resonance
point of spin-down (-up) electron under moderate magnetic field. In
this way, we can obtain both large electric current and large spin
current.

We now study the transport in T-stub waveguide with finite conductor and
sidearm widths. We
carry out
a numerical calculation for a waveguide
whose geometry parameters are $L=60a$, $N_x=10a$, $N_y=10a$.
The leads are assumed to have perfect Ohmic
contacts with the conductor. A hard wall potential is applied at the
edge of the waveguide.
This makes the lowest energy of
$n$-th subband (mode) in leads be
$\varepsilon_n=2|t_0|\{1-\cos[n\pi/(N_y+1)]\}$.
The Fermi energy in our calculation is between $0.083|t_0|$ and
$0.124|t_0|$ so that only the lowest mode contributes to the
conductance. The Zeeman splitting energy is $V_0=0.001|t_0|$
corresponding to a few Teslas for the typical III-V semiconductors
with lattice constance $a=20\AA$.
In Fig.~\ref{fig2}(b) we present the electron conductance as a
function of the injection electron energy
for a device with $L_s=60\,a$ and $L_m=15\,a$.
Noted that the energy is count from bottom of the first mode.
It is seen from the figure
that many properties of quasi-one-dimensional T-stub waveguide survive
in the finite width one: The conductance oscillates with the injection
energy. It approaches to zero at the anti-resonance points and then
quickly rise to about $e^2/h$ at the resonance points. And most
importantly, an anti-resonance point is alway accompanied by a
resonance point which is very close to it. In the presence
of modulation magnetic field, the difference of conductances between
these two spins is noticeably large when the inject energy locates in
one of the anti-resonance/resonance energy windows.
One can tune the Fermi Energy to obtain both large electric and spin
current. For example in the energy windows $[0.11619|t_0|,0.11789|t_0|]x$, we
obtain the largest spin current density with
$I_{\uparrow}^{SP}=I_{\uparrow}-I_{\downarrow}\approx5.986nA$ for the
spin-up current.

\begin{figure}[htb]
\vskip-0.3cm
\begin{center}
\psfig{figure=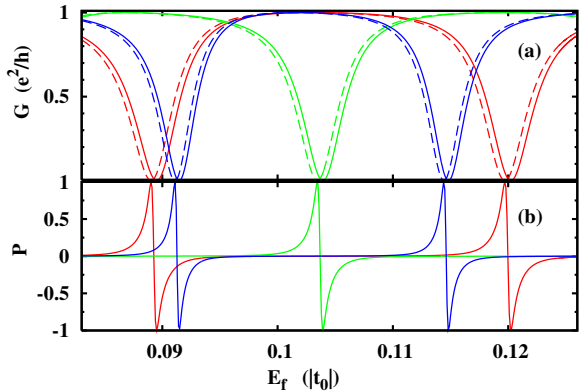,width=0.95\columnwidth}
\end{center}
\caption{(Color online) Spin dependent conductances $G$ (upper
panel) and the corresponding spin polarization as functions of the
electron energy for different sidearm length and fixed magnetic
modulation profile: Red curves: $L_s=60\,a$, $L_m=15\,a$; Green
curves: $L_s=75\,a$, $L_m=30\,a$; Blue curves: $L_s=90\,a$,
$L_m=45\,a$. The solid/dashed curves are conductances for spin-up/down electrons.}
\label{fig3}
\end{figure}

\begin{figure}[htb]
\vskip-0.3cm
\begin{center}
\psfig{figure=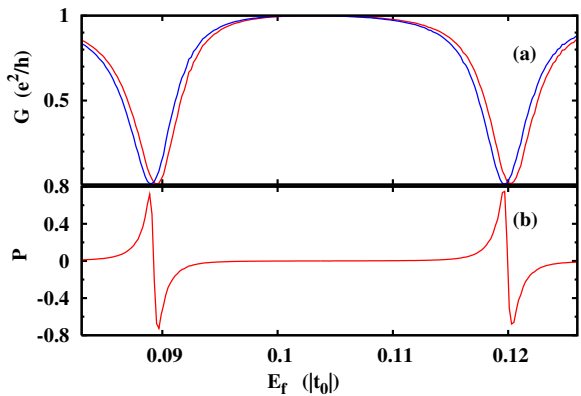,width=0.95\columnwidth}
\end{center}
\caption{(Color online)
The spin dependent conductance $G^{\sigma\sigma}$
(a) and SP (b) as functions of Fermi
energy for the device with Anderson disorder $W=0.005|t_0|$.
The blue and red curves in (a) correspond to $G^{\downarrow\downarrow}$ and
$G^{\uparrow\uparrow}$ respectively.
}
\label{fig4}
\end{figure}

Like the original QMT, we can also use the a remote gate voltage to
control the length of the sidearm. In Fig.~\ref{fig3}(b) we present
the conductances for different sidearm lengths $L_s$ but constant
$L_s-L_m$. In this way we simulate the electronically controlling of
the device without changing the magnetic modulation profile, {\it
i.e.} the strength and the positions of the applied magnetic field.
One can see from the figure that, the anti-resonance and the
corresponding resonance points change with the length of the sidearm
as they should have been. For a specified injection energy, when the
length of the sidearm changes, not only the conductance but also the
SP of output current change. For examples, in the energy window near
$0.0871|t_0|$ one gets about $100\%$ polarized spin-down current
when the sidearm is $60\,a$ long. Once the length of sidearm is
adjusted to $75\,a$, one gets about $8\%$ polarized current but the
direction of spin change to up. If one further adjust the length to
$90\,a$, the output current is almost non-polarized. It is seen that
with this filter one can control the strength, direction and
polarization rate of spin current electronically.


In order to further check the robustness of the spin filter we
propose, we now add Anderson disorder to the system and study its
effect on the SP. In Fig.~\ref{fig4}, the spin-dependent conductances
$G^{\uparrow \uparrow}$ and $G^{\downarrow \downarrow}$ as well as the
SP are plotted against the energy of the incident electrons with the
Anderson disorder included. The strength of the disorder
$W=0.005|t_0|$, five times of the modulated potential $V_0$. It is
found that the disorder decreases the transmission coefficients, but
only have slight effect on SP. In some energy windows, one can obtain
SP as high as $80\%$. Therefore the scheme we propose is robust
against the disorder.

In summary, we propose a spin filter scheme which enables the
electrically and magnetically remote control the spin polarization of
output current. The spin filter is a T-stub waveguide with a
modulation magnetic field at the tip of the sidearm. In this device,
electron conductance drops to the minimum when Fermi energy locates at the
anti-resonance points and rises to the maximum at resonance points.
With the modulation field, the (anti-)resonance points of spin-up and
-down electrons shift apart. Since a pair of the anti-resonance and
resonance points are very close to each other, one is able to use moderate
magnetic field to produce both large electric and spin currents.
Moreover one is able to control the direction and polarization
of the output spin current of the T-stub waveguide
via a remote gate which tunes the length of the sidearm and therefore
realize the remote electronically control of spin current.
We further shown that the device is robust against the
disorder.

The authors would like to thank M. W. Wu for proposing the topic
as well as directions during the investigation.  This work was supported by
the Natural Science Foundation of China
under Grant No.\ 10574120, 
the National Basic
Research Program of China under Grant No.\ 2006CB922005,
 the Knowledge Innovation
Project of Chinese Academy of Sciences and SRFDP. One of the authors (X.Y.F.)
thanks J. Zhou for valuable discussions.


\end{document}